\UseRawInputEncoding

\documentclass[a4paper, 10pt, conference]{ieeeconf}      

\IEEEoverridecommandlockouts                              

\usepackage{amsmath} 
\usepackage{amssymb}  
\usepackage[latin1]{inputenc}
\usepackage{here}
\usepackage{graphicx}
\usepackage{comment}
\usepackage{bm}

\title{\LARGE \bf
Dynamic network congestion pricing based on \\
deep reinforcement learning*
}

\author{Kimihiro Sato$^{1}$, Toru Seo$^{2}$ and Takashi Fuse$^{1}$
\thanks{*This work was mainly supported by a JSPS KAKENHI Grant-in-Aid for Scientific Research 20H02267.}
\thanks{$^{1}$K. Sato and T. Fuse are with Department of Civil Engineering, the University of Tokyo, Tokyo 113-8656, Japan.}%
\thanks{$^{2}$T. Seo is with the Department of Civil and Environmental Engineering, Tokyo Institute of Technology, Tokyo 152-8552, Japan. Email: {\tt\small seo.t.aa@m.titech.ac.jp}}%
}

\begin{document}

\maketitle
\thispagestyle{empty}
\pagestyle{empty}

\begin{abstract}
Traffic congestion is a serious problem in urban areas.
Dynamic congestion pricing is one of the useful schemes to eliminate traffic congestion in strategic scale.
However, in the reality, an optimal dynamic congestion pricing is very difficult or impossible to determine theoretically, because road networks are usually large and complicated, and behavior of road users is uncertain.
To account for this challenge, this work proposes a dynamic congestion pricing method using deep reinforcement learning (DRL).
It is designed to eliminate traffic congestion based on observable data in general large-scale road networks, by leveraging the data-driven nature of deep reinforcement learning.
One of the novel elements of the proposed method is the distributed and cooperative learning scheme.
Specifically, the DRL is implemented by a spatial-temporally distributed manner, and cooperation among DRL agents is established by novel techniques we call spatially shared reward and temporally switching learning.
It enables fast and computationally efficient learning in large-scale networks.
The numerical experiments using Sioux Falls Network showed that the proposed method works well thanks to the novel learning scheme.
\end{abstract}

\section{INTRODUCTION}

Traffic congestion is a serious problem in urban areas.
It damages the economy due to the wasted travel time, as well as traffic safety and environment.
Appropriate control schemes that reduce or eliminate congestion are required.

{\it Dynamic congestion pricing} \cite{vickrey} is one of the useful schemes to eliminate traffic congestion in strategic scale.
Congestion is mainly caused by spatial and/or temporal concentration of traffic demand.
In a dynamic congestion pricing scheme, appropriate tolls are charged to road users in certain roads in certain time duration to mitigate the demand concentration.
In the literature, it has been theoretically shown that an optimal dynamic congestion pricing could eliminate traffic congestion, if the road network shape is simple and the behavioral principle of road users is known \cite{vickrey}.

In the reality, an optimal dynamic congestion pricing is very difficult or impossible to determine theoretically, because road networks are usually large and complicated, and behavior of road users is uncertain.
To account for this challenge, {\it trial-and-error pricing schemes} have been proposed \cite{Yang2004} \cite{inproceedings} \cite{SATO2021347}. 
They charge a trial toll first, and then adjust a toll in the next term based on the road user's response to the trial toll.
By repeating this trial-and-error process for a certain duration, they eventually find the optimal toll.
Unfortunately, this process sometimes takes a long time (e.g., more than a year) in large-scale networks, which means that conventional trial-and-error method may be unpractical in such situations.

The {\it deep reinforcement learning (DRL)} can be considered as an efficient data-driven approach to execute such trial-and-error process \cite{Pandey2020}. 
The notable advantages of DRL are as follows.
First, it can automatically extract meaningful patterns from massive observation data.
Second, it can automatically optimize the adjustment process using the data.
These advantages are useful for dynamic congestion pricing optimization in a large road network.

The goal of this work is to develop a dynamic congestion pricing method using DRL.
It determines efficient congestion tolls based only on observation data (e.g., travel time) in general large-scale networks in a day-to-day setting.
We consider a general transportation model in which travelers can choose their departure time and routes simultaneously.
To achieve the goal, this work develops a novel distributed and cooperative learning scheme for DRL is developed.
Specifically, the DRL is implemented by a spatial-temporally distributed manner, and cooperation among DRL agents is established by novel techniques we call {\it spatially shared reward} and {\it temporally switching learning}.
Thanks to this scheme, the method's learning speed is computationally efficient and fast, even in large-scale applications.
The proposed method is validated using actual road network data and compared with other methods, and the effectiveness of the proposed learning scheme was confirmed.

\section{LITERATURE REVIEW}

Application of reinforcement learning (RL) to transportation problems can be reviewed as follows.
Wang et al. \cite{e21080744} apply deep reinforcement learning to traffic signal control. Their proposed method uses "high-resolution event-based data" , which keep track of passage and presence of vehicles by recording activation/deactivation events of vehicle detectors.
Li et al. \cite{LI2021103059} proposed "Knowledge Sharing Deep Deterministic Policy Gradient (KS-DDPG)", which introduces a communication protocol for knowledge sharing among multi-agents, and applies it to cooperative control of traffic signals.
Li et al. \cite{LiStrategy2020} propose a DDPG-based driving strategy for individual vehicles to mitigate oscillations in stop-and-go-waves and optimize traffic safety. It is shown to reduce collision risk.


Some studies use RL or neural networks to determine dynamic congestion tolls.
Zhu and Ukkusuri \cite{zhu} apply RL to dynamic congestion pricing with distance-based tolls. Toll is got closer to the optimal one by progressing a learning process with traffic flow data. 
Mirzaei et al. \cite{Mirzaei8569737} apply RL to $\Delta$-tolling \cite{PartC17-Sharon}, which updates tolls in response to the difference between the current travel time on each link and its free flow travel time.
Pandey et al. \cite{Pandey2020} apply DRL to congestion charging in a traffic model with a mixture of managed and general lanes.
Genser and Kouvelas \cite{GENSER2022103485} propose to use multilayer neural networks to determine dynamic congestion tolls for inter-area travel.
Regardless, to our knowledge, DRL methods have not been utilized to determine dynamic congestion tolls in general road networks with day-to-day departure time and route choice problem.

\section{TRAFFIC MODEL}

The definition of the traffic model considered in this study is as follows.
We consider a dynamic traffic model where departure time and route are determined by day-to-day dynamics. 
The road network of the model has multiple Origin Destination (OD) pairs connected by multiple routes.
Each link may have a bottleneck.
Origins are residential areas, and destinations are central business districts (CBD).
The dynamic traffic model is an extension of the bottleneck model \cite{vickrey}. \par

Let $j$ be day, $t$ be time, $t^{\ast}$ be the travelers' desired arrival time to the destination. Let $\mu_i$ be the capacity of bottleneck $i$ per unit time, $a_{j,i}(t)$ be the inflow rate at bottleneck $i$ on day $j$, $N_{j,i}(t)$ be the number of vehicles in waiting queue on day $j$ and at bottleneck $i$, $w_{j,i}(t)$ be waiting time on day $j$ and at bottleneck $i$, and $\tau_{j,i}(t)$ be tolls on day $j$ and at bottleneck $i$. $t$ is time when travelers leave bottleneck $i$ in them. In addition, let $B_z$ be a given set of all bottlenecks on route $z$, and $Z_{\zeta}$ be a given set of all routes connecting OD pair $\zeta$. Waiting queues are point queues with no physical length.

The queue evolution at time $t$ is defined as Eq. (\ref{que}).
\begin{equation}
	\label{que}
	\frac{\mathrm{d}N_{j,i}(t)}{\mathrm{d}t}=\left\{\begin{array} {ll}
		0\qquad (N_{j,i}(t)=0\ \mbox{and}\ a_{j,i}(t)<\mu_i) \\
		a_{j,i}(t)-\mu_i \hspace{55.5pt} (\mbox{otherwise}) 
	\end{array} \right.
\end{equation}

The waiting time in the queue at time $t$ is defined as Eq. (\ref{wt}).
\begin{equation}
	\label{wt}
	w_{j,i}(t)=\frac{N_{j,i}(t)}{\mu_i}
\end{equation}

The generalized cost of a traveler who arrives at the destination at time $t$ through route $z$ on day $j$ is defined as
\begin{equation}
	\label{gcost}
	\begin{split}
	&c_{j,z}(t)=\left\{\begin{array} {r}
		\sum\limits_{i\in B_z}\tau_{j,i}(t_{\mathrm{l},z,i}(t))+\alpha \{t-t_{\mathrm{d},z}(t)\}+\beta\left(t^\ast-t\right) \\
		(t<t^{\ast})\\
		\\
		\sum\limits_{i\in B_z}\tau_{j,i}(t_{\mathrm{l},z,i}(t))+\alpha \{t-t_{\mathrm{d},z}(t)\}+\gamma\left(t-t^\ast\right) \\
		(t\geq t^{\ast})
	\end{array} \right.
	\end{split}
\end{equation}
where $t_{\mathrm{d},z}(t)$ is a traveler's departure time from an origin when one arrives at a destination at time $t$ through route $z$, 
$t_{\mathrm{l},z,i}(t)$ is a traveler's departure time from bottleneck $i$ when one arrives at a destination at time $t$ through route $z$, 
$\alpha$ is the value of time per unit time for a traveler, $\beta$ is the early arrival penalty per unit time, and $\gamma$ is the late arrival penalty per unit time.

Multinominal logit model \cite{YU202059} is used for day-to-day dynamics.
Travel cost is defined with the weighted average learning operator \cite{CASCETTA19891} \cite{OUYANG} as "perceived travel cost". Note that its definition is extended in this study.
The probability of choosing departure time $t$ and route $z$ between OD pair $\zeta$ is defined as
\begin{equation}
	\label{mlogit}
	P_{j,z}^{\zeta}(t)=\frac{\mathrm{exp}(-\vartheta\overline{c}_{j,z}(t))}{\sum\limits_{(z',t')}\mathrm{exp}(-\vartheta\overline{c}_{j,z'}(t'))}
\end{equation}
where $\vartheta$ is a parameter, and $\overline{c}_{j,z}(t)$ is the perceived travel cost at day $j$, time $t$, and route $z$.
Perceived travel cost is defined as
\begin{equation}
	\begin{split}
		\label{pcost}
		\overline{c}_{j,z}(t)=\frac{1}{\varsigma(\lambda)}\left(c_{j-1,z}(t)+\sum\limits_{i=2}^{T_c}\lambda^{i-1}c_{j-i,z}(t)\right) \\
		\forall z\in Z_{\zeta}
	\end{split}
\end{equation}
\begin{equation}
	\label{sl}
	\varsigma(\lambda)=\left\{
	\begin{array}{ll}
		\sum_{i=1}^{T_c}\lambda^{i-1} & (0<\lambda <1) \\
		1 & (\lambda=0)
	\end{array} \right.
\end{equation}
where $\lambda$ and $T_{\mathrm{c}}$ is a parameter, and $1/\varsigma(\lambda)$ is a normalization factor.

Number of departing vehicles by departure time and route is defined as
\begin{equation}
	\label{depvol}
	f_{j,z}(t)=M_{\zeta}\cdot P_{j,z}^{\zeta}(t)\qquad \forall z\in Z_{\zeta}
\end{equation}
where $f_{j,z}(t)$ is the number of departing vehicles by departure time and route, and $M_{\zeta}$ is traffic demand between OD pair $\zeta$ per day.

Bounded rationality \cite{YU202059} is used with day-to-day dynamics. It means that the departure time and route choice will not be changed on the next day when the difference between the expected cost of the current choice and the minimum expected cost of all other choices is less than or equal to a threshold $\delta$.
When a traveler who chooses route $z$ and departure time $t$ on day $j-1$ does not change departure time and route choice on day $j$, the expected cost $\widehat{c}_{j,z}(t)$ satisfies Eq. (\ref{BR_1}).
\begin{equation}
	\label{BR_1}
	\widehat{c}_{j,z}(t)-\widehat{c}_{j,z'}(t')\leq \delta \qquad \forall(z',t')\neq (z,t)
\end{equation}

\section{CONGESTION PRICING METHOD}

\subsection{Overview}

\begin{figure}[tb]
	\centering
	\includegraphics[width=0.9\linewidth]{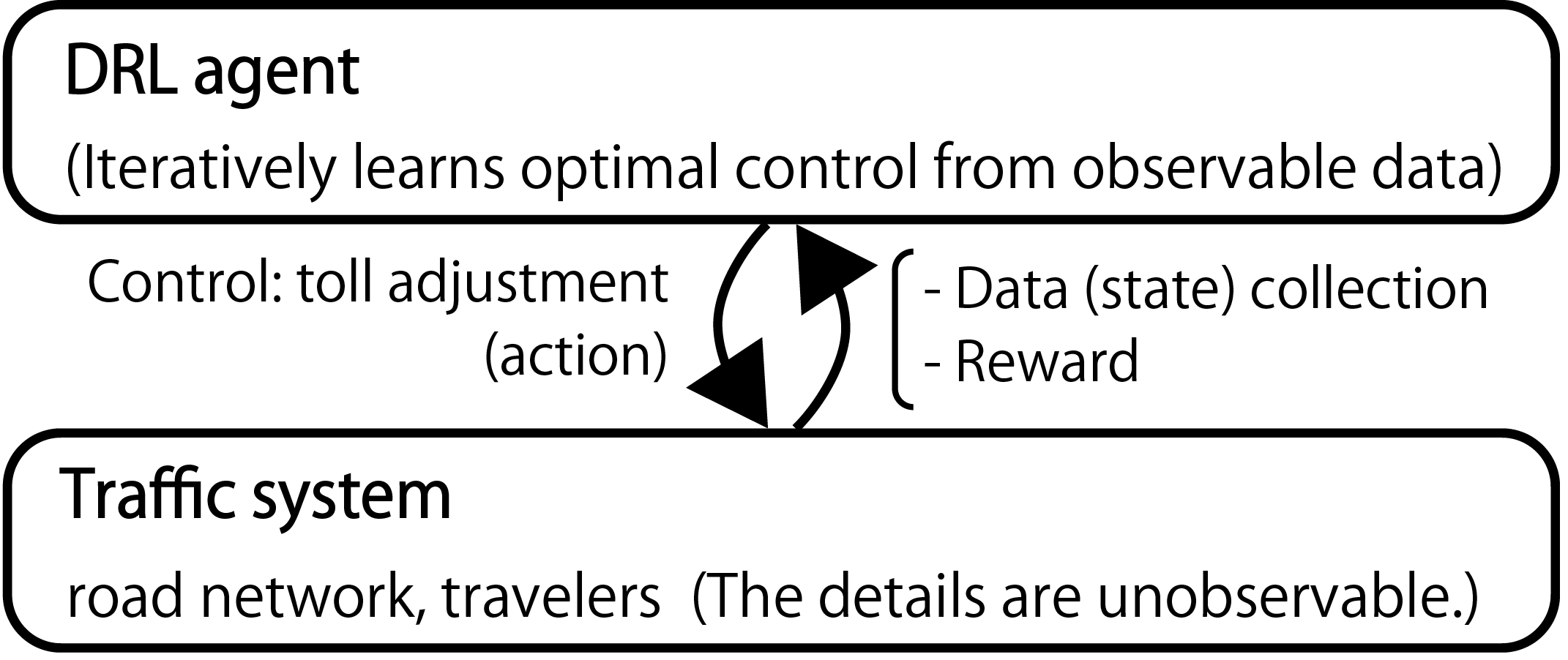}
	\caption[The outline of DRL]{The outline of DRL}
	\label{DRL}
\end{figure}

In this study, we propose "Distributed Pricing Deep Deterministic Policy Gradient (DP-DDPG)", which is a DRL-based dynamic congestion pricing method.
Fig. \ref{DRL} shows the outline of DRL.
In DP-DDPG, traffic data and toll update by bottleneck and time slot are handled in DRL. In other words, a DRL agent performs "distributed control". It significantly reduces the dimensionality of the states and actions.
Specifically, an DRL agent observes traffic data as a state at each pair of a bottleneck and a time slot, and learns toll adjustment as an action at each pair of a bottleneck and a time slot with reward.

In distributed control, cooperation among pairs of a bottleneck and a time slot is required. In this study, we propose two methods for the cooperation: spatially shared reward and temporally switching learning.

Deep Deterministic Policy Gradient (DDPG) \cite{lillicrap2015continuous} is used as the framework for DRL.
DDPG is one of DRL algorithms which deal with actions as continuous values, uniquely decides an action for a state (deterministic policy), and is based on actor-critic. 
Due to the space limitation, the detailed formulation and algorithm of DDPG are omitted from this paper; interested readers may refer the original paper \cite{lillicrap2015continuous}.

\begin{figure*}[tb]
	\centering
	\includegraphics[width=\linewidth]{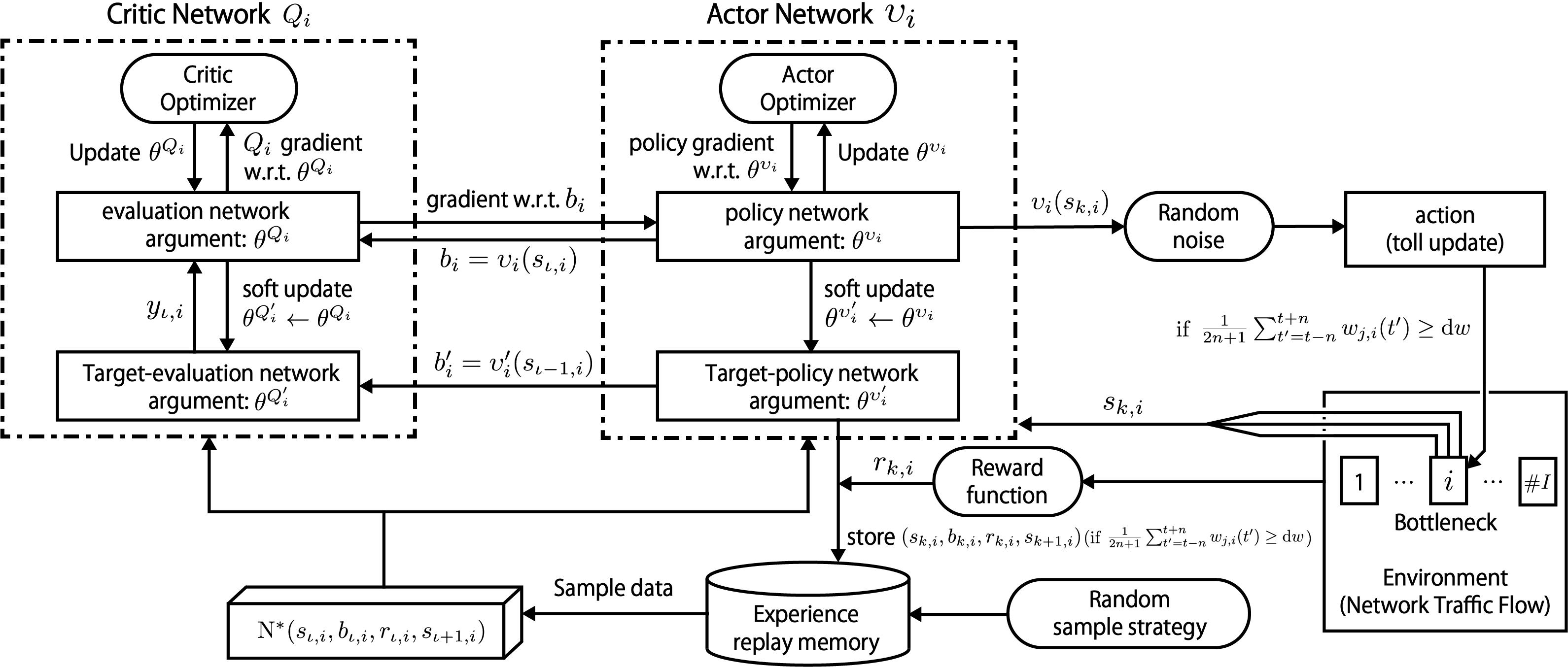}
	\caption[The flow of DP-DDPG]{The flow of DP-DDPG}
	\label{DDPG_OL}
\end{figure*}

The flow of DP-DDPG is shown as Fig. \ref{DDPG_OL}. Guo et al. \cite{guo2020autonomous} is used as its reference.

\subsection{Definition of state, action, and reward}

A state is based from inflow rate to a bottleneck, waiting time and a toll.
The state at day $j$, time $t$, and bottleneck $i$ is defined as
\begin{equation}
	\label{state_form}
	\begin{split}
		\bm{s}_{j,i}(t)=&\left( \frac{a_{j,i}(t)-\mu_i}{\mu_i}, \frac{w_{j,i}(t)}{\sum\limits_{t\in T}w_{0,i}(t)/\sum\limits_{t\in T}\varpi_{0,i,t}},\right. \\
		&\left. \quad \frac{\tau_{j,i}(t)-\bar{\tau}_{j,i}}{\sum\limits_{t\in T}w_{0,i}(t)/\sum\limits_{t\in T}\varpi_{0,i,t}} \right)
	\end{split}
\end{equation}

\begin{equation}
	\label{sub_state_form}
	\varpi_{j,i,t}=\left\{
	\begin{array}{ll}
		1 & (w_{j,i}(t)>0) \\
		0 & (w_{j,i}(t)=0)
	\end{array}
	\right.
\end{equation}
where $t$ is time when travelers leave bottleneck $i$, and $\bm{s}_{j,i}(t)$ is the state at day $j$, time $t$, and bottleneck $i$.

In order to input the state to neural networks, the scale of each element is aligned by a capacity of a bottleneck and waiting time when no toll is charged and day-to-day dynamics converge.

An action is a range of increase/decrease in a toll.
The action at day $j$, time $t$, and bottleneck $i$ is defined as
\begin{equation}
	\label{action_output}
	b_{j,i}(t)=G\cdot \tanh y\quad (-G<b_{j,i}(t)<G)
\end{equation}
where $b_{j,i}(t)$ is the action at day $j$, time $t$, and bottleneck $i$, $y$ is an input to the final layer of an actor network, and $G$ is a parameter which decides the range of an action.

The toll at day $j$, time $t$, and bottleneck $i$ is updated as Eq. (\ref{toll_update}).
\begin{equation}
	\label{toll_update}
	\tau_{j,i}(t)\leftarrow \tau_{j,i}(t)+b_{j,i}(t)
\end{equation}

A reward is constructed from the below two term.

\begin{itemize}
	\item a term based on waiting time by time slot and bottleneck
	\item a term based on the average waiting time at all bottlenecks where tolls are charged and all time slots (spatially shared reward)
\end{itemize}

The purpose of using spatially shared reward is cooperative decrease among all bottlenecks where tolls are charged and all time slots.
In a general road network, fluctuation in waiting time at one bottleneck can affect that at other bottlenecks.
The term is expected to prevent from toll adjustment which increases waiting time at other bottlenecks.

The reward at day $j$, time $t$, and bottleneck $i$ is defined as
\begin{equation}
	\label{rew}
	\begin{split}
		r_{j,i}(t)=&-\left\{ \frac{w_{j,i}(t)}{\sum\limits_{t\in T}w_{0,i}(t)/\sum\limits_{t\in T}\varpi_{0,i,t}}\right. \\
		&\left. \qquad +\frac{1}{\# I}\sum_{i\in I}\frac{\bar{w}_{j,i}}{\sum\limits_{t\in T}w_{0,i}(t)/\sum\limits_{t\in T}\varpi_{0,i,t}}
		\right\}
	\end{split}
\end{equation}
where $r_{j,i}(t)$ is the reward at day $j$, time $t$, and bottleneck $i$, and $I$ is a set of bottlenecks where tolls are charged.

\subsection{Temporally switching learning}

In this study, we propose "temporally switching learning". It means that an agent does not perform DRL and fixes its action at zero in the pair of a time slot $t'$ and a bottleneck $i$ where a time slot moving average of waiting time is less than a threshold $\delta$, and it is defined as
\begin{equation}
	\label{kufu}
	\frac{1}{2n+1} \sum_{t'=t-n}^{t+n}w_{j,i}(t')<\mathrm{d}w
\end{equation}
where $n\in \mathbb{N}$ is a parameter, and $\mathrm{d}w$ is a constant.

The first purpose of temporally switching learning is improvement of the efficiency of DRL. It narrows down the learning target of an agent to toll adjustment in the pair of a time slot and a bottleneck where waiting time is not nearly zero.

Its second purpose is prevention against excessive increase of tolls and destabilization of learning in the pair of a time slot and a bottleneck where DRL is turned off. 
In the pair of a time slot and a bottleneck where waiting time keeps nearly zero, relatively high reward is given to an agent. This may cause continuous increase of a toll and make learning unstable.

\section{NUMERICAL EXPERIMENTS}

\subsection{Purpose and scenario}

The performance of DP-DDPG is evaluated by various types of numerical experiments. The summary of the experiment scenarios and their purposes are as follows:
\begin{itemize}
	\item DP-DDPG is applied to Sioux Falls Network to evaluate the performance in the condition where multiple OD pairs and multiple routes exist.
	\item DP-DDPG is applied to a parallel bottleneck model, where route choice is important, to compare it with other methods.
\end{itemize}

\subsection{Sioux Falls Network}
\begin{figure}[tb]
	\centering
	\includegraphics[width=\linewidth]{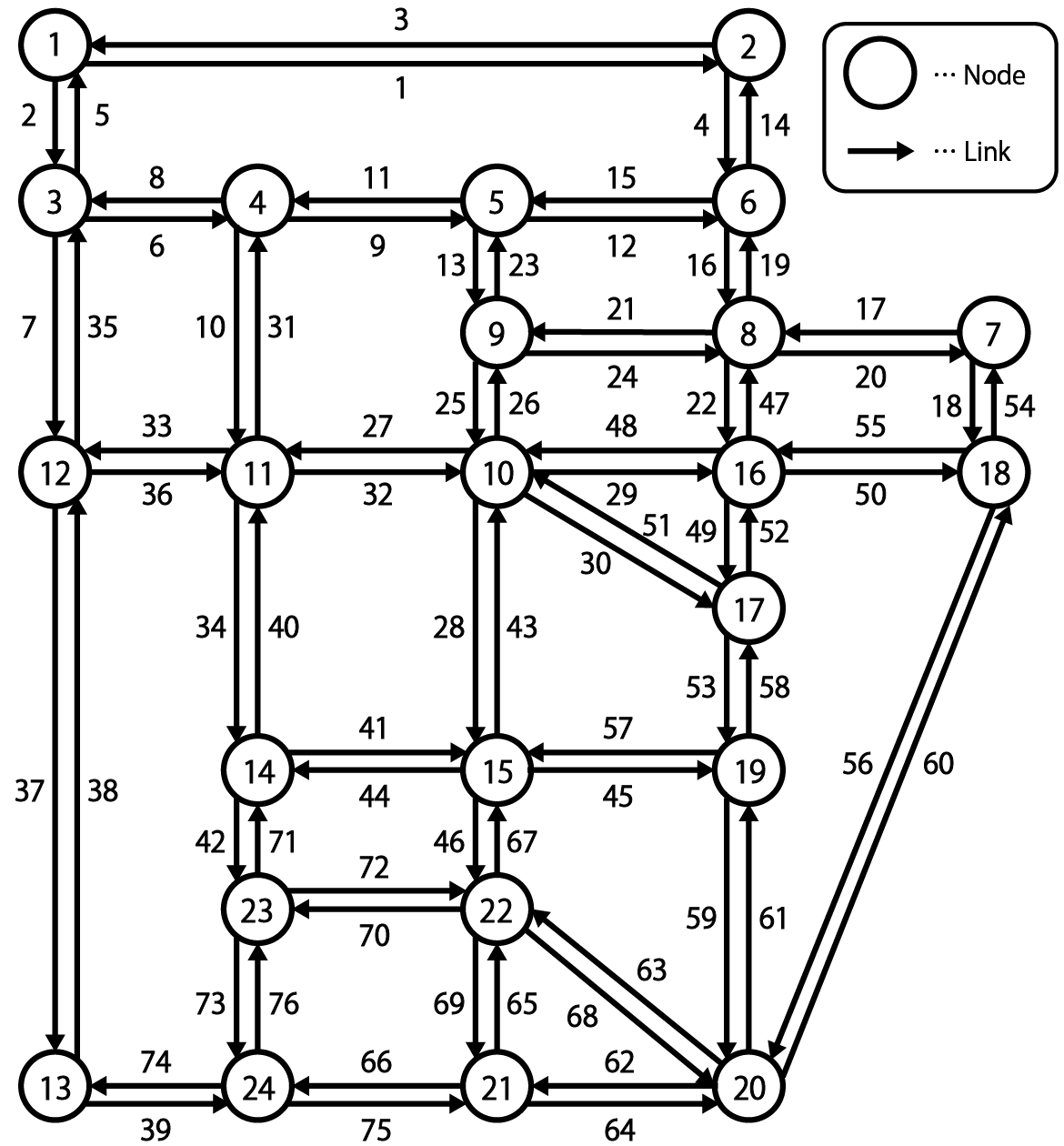}
	\caption[Sioux Falls Network]{Sioux Falls Network}
	\label{SFnet}
\end{figure}

We used Sioux Falls Network \cite{northwestern1973development} as illustrated in Fig. \ref{SFnet}.
Numbers of links with a bottleneck where a toll is charged and waiting time is not zero are 29, 48, 53, and 58.
Those with a bottleneck where a toll is not charged and waiting time is not zero are 49, 52, and 61.
$t^\ast$ is set to be equal for all travelers. 

DRL is performed in several patterns of values of learning rate of actor networks and critic networks, and $G$. We decided that optimal parameters as follows:
\begin{itemize}
	\item learning rate of actor networks: $10^{-3}$
	\item learning rate of critic networks: $10^{-2}$
	\item $G$: $1.5$
\end{itemize}
The parameters of traffic model are specified as follows:
$\alpha=1.0$, $\beta=0.45$, $\gamma=1.2$, $\vartheta=0.015$, $\lambda=0$, $\delta=1$, $t\in\mathbb{N}$, $1\leq t\leq 250$, and $t^\ast=75$.
During DRL, 40 days are set as one cycle, and after 40 days have passed, DRL is returned to the initial state. It is when "$\tau_{j,i}(t)=0\hspace{10pt} \forall (j,i,t)$ and day-to-day dynamics converge". Values of the parameters of actor networks and critic networks are succeeded. One set of DRL consists of 15 cycles, and 10 sets are performed with the above values of learning rates and $G$.

\begin{figure}[tb]
	\centering
	\includegraphics[width=\linewidth]{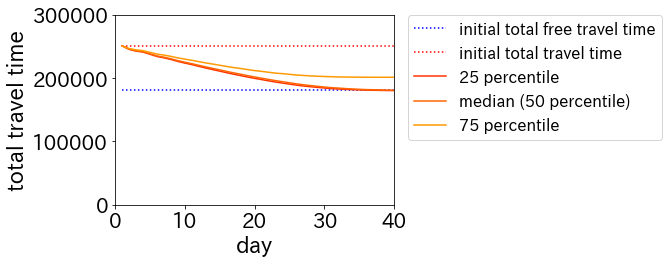}
	\caption[Day-to-day dynamics of total travel time in Sioux Falls Network]{Day-to-day dynamics of total travel time in Sioux Falls Network}
	\label{totaltt}
\end{figure}

Fig. \ref{totaltt} shows the transition of total travel time in Sioux Falls Network when the trained agents in each set are applied to it under the same conditions.
On Fig. \ref{totaltt}, initial total free travel time and initial total waiting time are those when $\tau_{j,i}(t)=0\hspace{10pt} \forall (j,i,t)$ and day-to-day dynamics converge.
According to Fig. \ref{totaltt}, DP-DDPG stably reduces total travel time in about 40 days.

\begin{figure*}[tb]
	\centering
	\includegraphics[width=\linewidth]{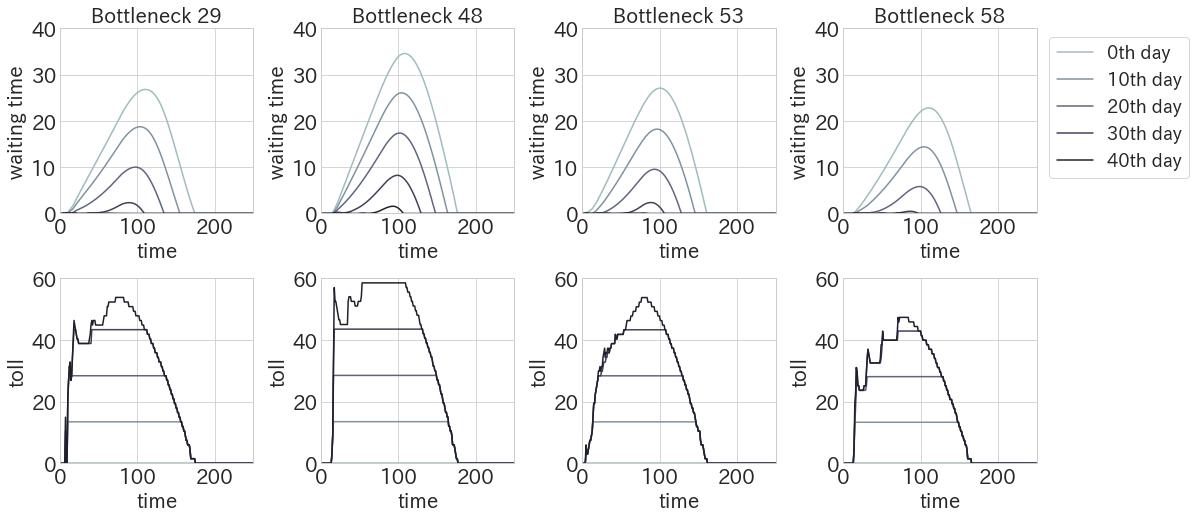}
	\caption[Transition of waiting time distribution and toll distribution]{Day-to-day dynamics of within-day waiting time (up) and toll (down) distribution in each bottleneck}
	\label{dist}
\end{figure*}

Fig. \ref{dist} shows waiting time distribution and toll distribution at each bottleneck.
According to Fig. \ref{dist}, tolls increase according to waiting time distributions.

\subsection{Parallel bottleneck model}

\begin{figure}[tb]
	\centering
	\includegraphics[width=\linewidth]{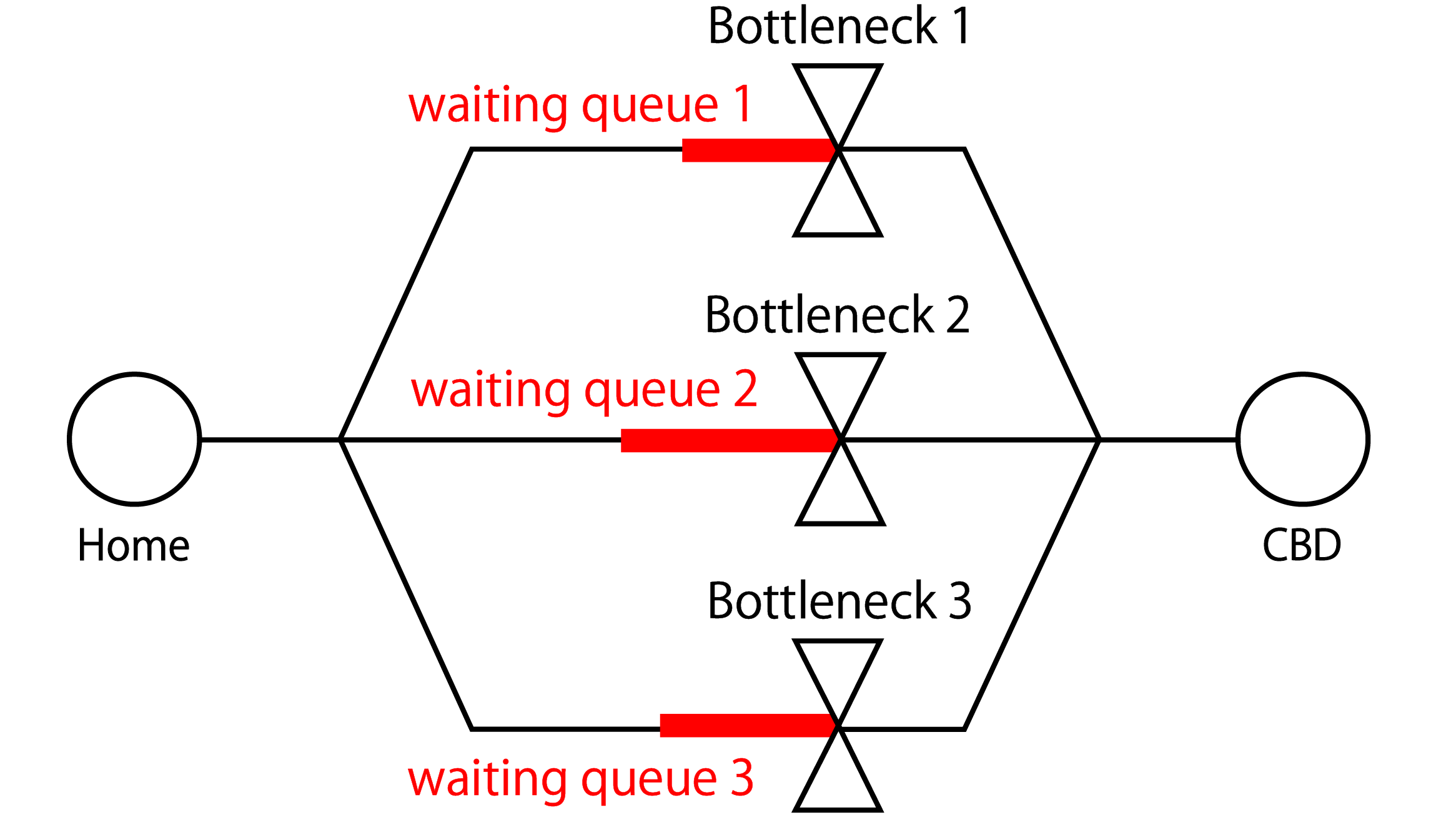}
	\caption[Illustration of the parallel bottleneck model \cite{SATO2021347}]{Illustration of the parallel bottleneck model \cite{SATO2021347}}
	\label{BN_three}
\end{figure}

We used a parallel bottleneck model with a single OD pair connected three path and one bottleneck per path as illustrated in Fig. \ref{BN_three}. It is the same model as "multiple bottleneck model" \cite{SATO2021347}.
$t^\ast$ is set to be equal for all travelers.

We compared the result of DP-DDPG with those of centralized DDPG, fully distributed DDPG and Q-learning \cite{SATO2021347}. 
The outline of each method is as follows.
In centralized DDPG, a state is defined with traffic data at all bottlenecks where tolls are charged, an action is defined with toll update there, and a reward is based on waiting time there.
Toll distributions are set as piecewise linear functions.
In other methods, a state is defined with traffic data at each bottleneck where a toll is charged, and an action is defined with toll update there.
In DP-DDPG (our proposed method) and fully distributed DDPG, a reward is based on waiting time at each bottleneck where a toll is charged. 
In Q-learning, a reward is based on inflow rate at each bottleneck where a toll is charged. 
In DP-DDPG and Q-learning, spatially shared reward is incorporated.

Values of learning rate of actor networks and critic networks, and $G$ are as follows:
\begin{itemize}
	\item learning rate of actor networks: $10^{-5}$
	\item learning rate of critic networks: $10^{-4}$
	\item $G$: $0.5$
\end{itemize}
The parameters of traffic model are specified as follows:
$\alpha=1.0$, $\beta=0.45$, $\gamma=1.2$, $\vartheta=0.05$, $\lambda=0$, $\delta=0.5$, $t\in\mathbb{N}$, $1\leq t\leq 80$, and $t^\ast=30$.

During DRL by methods except Q-learning, 60 days are set as one cycle, and after 60 days have passed, DRL is returned to the initial state. It is when "$\tau_{j,i}(t)=0\hspace{10pt} \forall (j,i,t)$ and day-to-day dynamics converge". Values of the parameters of actor networks and critic networks are succeeded. One set of DRL consists of 20 cycles, and 20 sets are performed with the above values of learning rates and $G$.

During RL by Q-learning, 1000 days are set as one cycle, and after 1000 days have passed, RL is returned to the initial state. It is when "$\tau_{j,i}(t)=0\hspace{10pt} \forall (j,i,t)$ and day-to-day dynamics converge". The update result of parameter is succeeded. One set of RL consists of 30 cycles, and three sets are performed. 
Note that replicator dynamics is used as day-to-day dynamics in Q-learning.

\begin{figure*}[tb]
	\centering
	\includegraphics[width=\linewidth]{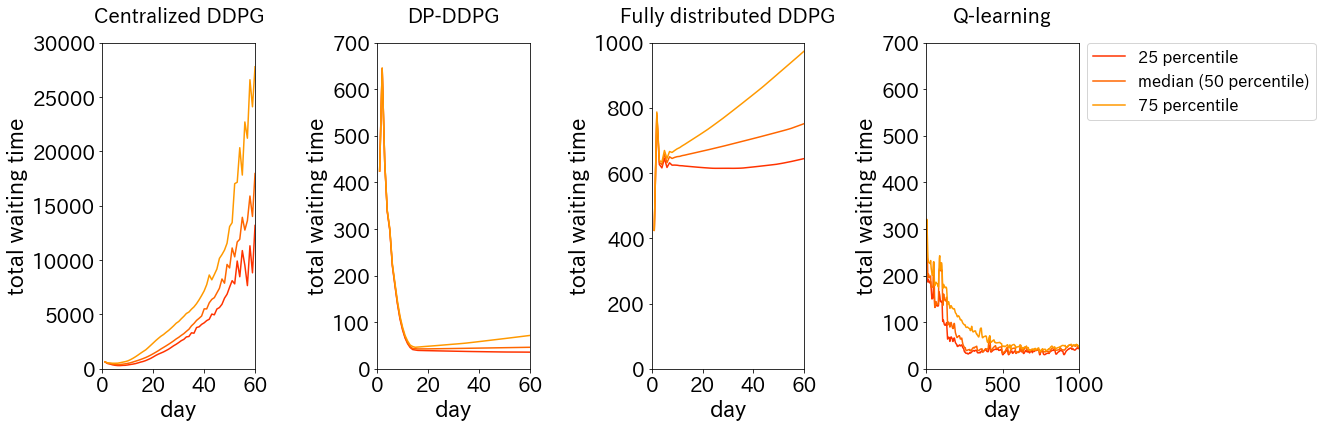}
	\caption[Day-to-day dynamics of total waiting time in each method]{Day-to-day dynamics of total waiting time in each method}
	\label{comp_ALL}
\end{figure*}

Fig. \ref{comp_ALL} shows the transition of total waiting time in all methods.
According to Fig. \ref{comp_ALL}, learned agents by centralized DDPG and those by fully distributed DDPG could not decrease waiting time.
In centralized DDPG, since we used piecewise linear functions as toll distributions, they could not be made close to the proper toll distributions, and day-to-day dynamics could not become stable.
In fully distributed DDPG, since cooperation among pairs of a bottleneck and a time slot is not used, agents could not learn properly in the condition where multiple bottlenecks exist.
According to Fig. \ref{comp_ALL}, learned agents by Q-learning can decrease waiting time. It can be because spatially shared reward was used. However, once waiting time became nearly zero, it oscillated.
It can be because actions are discrete values and temporally switching learning is not used. 

Note that waiting time reduction speed is not generally comparable because replicator dynamics is used for the day-to-day dynamics and tolls are updated every time day-to-day dynamics converge.

\section{CONCLUSION}

A dynamic congestion pricing method using deep reinforcement learning is proposed.
It is designed to eliminate traffic congestion based on observable data in general large-scale road networks.
One of the novel elements of the proposed method is the distributed and cooperative learning scheme.
It enables a fast and computationally efficient learning in large-scale networks.
The numerical experiments using Sioux Falls Network showed that the method works well thanks to the novel learning scheme.

\addtolength{\textheight}{-12cm}   

\end{document}